\documentstyle[aps,prl,psfig]{revtex}

\begin{document}
\draft
\twocolumn[\hsize\textwidth\columnwidth\hsize\csname
@twocolumnfalse\endcsname
\title{$1/f^\alpha$ noise from correlations between
avalanches in self-organized criticality}
\author{J\"orn Davidsen$^{1,2}$ \cite{byline} and Maya Paczuski$^1$}
\address{
$^1$ Department of Mathematics, Imperial College of Science, Technology,
and Medicine, London, UK SW7 2BZ\\
$^2$ Institut f\"ur Theoretische Physik und Astrophysik,
Christian-Albrechts-Universit\"at, Olshausenstra\ss e 40, 24118 Kiel,
Germany}
\date{\today}
\maketitle

\begin{abstract}
We show that large, slowly driven systems can evolve to a
self-organized critical state where long range temporal
correlations between bursts or avalanches
produce low frequency $1/f^{\alpha}$ noise.  The
avalanches can occur instantaneously in the external time scale of the
slow drive, and their event statistics are described by power law
distributions.  A specific example of this behavior is provided by
numerical simulations of a deterministic ``sandpile''  model.

\end{abstract}
% insert suggested PACS numbers in braces on next line
]

\narrowtext

The ubiquity of $1/f^\alpha$ noise in nature is one of the oldest
problems in contemporary physics still lacking a generally accepted
explanation, despite much effort.  The phenomenon is characterized by
a $1/f^\alpha$ decay with a nontrivial $\alpha \approx 1$ found in the
power spectrum $S(f)$ of a given time signal at low frequencies
$f$. It has been observed in a huge number of diverse systems, many of
which are far from equilibrium and show an avalanche-like dynamics.
Examples include earthquakes \cite{davidsen01a,sornette}, combustion
fronts \cite{skokov01}, chemical reactions \cite{claycomb01}, flux
motion in superconductors \cite{field95}, and Barkhausen noise
\cite{tadic99}, to name only a few.

In the search for a general dynamical mechanism for $1/f^\alpha$ noise
in far from equilibrium systems, Bak, Tang and Wiesenfeld (BTW)
proposed the concept of self-organized criticality (SOC)
\cite{bak87,bak}.  This refers to the tendency of spatially extended,
slowly driven systems to organize into a state with fractal spatial
and temporal properties that is also characterized by self-similar
distributions of event (avalanche) sizes.  Although the original
``sandpile'' model, which was introduced to exemplify the SOC concept,
exhibits a scale-free distribution of avalanches, its noise spectrum
is of the form $1/f^2$ \cite{cjf}.  Variants of the original sandpile 
model do indeed exhibit nontrivial $1/f^{\alpha}$ noise, but in these 
cases \cite{maslov99} the avalanche event distributions are not 
critical, implying, possibly, mutual exclusivity between a SOC mechanism 
and mechanisms giving long range temporal correlations such as
$1/f^{\alpha}$ noise.  

In fact, it has recently been argued in the
context of solar flares, transport dynamics in magnetic confinement
devices, and other areas, that the presence of temporal
correlations between events excludes SOC as an underlying mechanism.
For instance, in Ref. \cite{carbone} it was stated that for SOC ``one
expects no correlation between ...  bursts''.  See Ref. \cite{solar}
for a discussion and references.  The argument entails a narrowing of
SOC to the phenomenology of certain, specific ``sandpile'' models,
which is in sharp contrast to the original idea \cite{bak87,bak}, and
is erroneous as we show here.

In order to demonstrate explicitly that SOC can  provide
a dynamical mechanism which gives correlations between bursts leading
to $1/f^{\alpha}$ noise, we study a slowly driven,
 deterministic ``sandpile'' model.
It exhibits a power-law distribution of avalanches, as well as
$1/f^{\alpha}$ fluctuations in the pattern of dissipation over the
slow temporal domain of the external drive.  Despite the fact that the
time scales of the driving and of the avalanche events are completely
separated, the critical behavior of the power spectrum is solely
determined by the critical properties of the avalanche size
distribution.  In addition, the results we find are robust with
respect to changes in the definition of the time scale associated with
the driving.  These observations  constitute a
proof that a SOC mechanism can give low frequency $1/f^{\alpha}$ noise
due to correlations between power-law distributed avalanches, without
imposing  temporal correlations in the external drive.

Actually several different models of SOC, describing e.g.  traffic
\cite{nagel95} and evolution \cite{bak,paczuski96} do exhibit
nontrivial $1/f^{\alpha}$ noise over the temporal domain of individual
avalanches.  In this context, it is important to distinguish between
temporal correlations within the time span of individual avalanche
events, and low frequency noise observed over a much longer temporal
regime of an arbitrarily slow external drive. Measured in this
external time scale, the individual avalanches can occur almost
instantaneously.  This is a situation often encountered in Nature.  In
this case $1/f^{\alpha}$ noise must arise from correlations between
avalanches.  These correlations can either be induced by long range
temporal correlations in the external drive, or be an intrinsic part
of the self-organization process itself. Sanchez {\it et al}
 have found long range temporal correlations in SOC when the
``sandpile'' is driven by an external source of low frequency noise
\cite{solar}.

However, it is important to clarify if SOC itself can spontaneously
generate both critical avalanche statistics and long-range temporal
correlations between avalanches, in the presence of a temporally
uniform, slow external drive.  Surprisingly, this question has not yet
been decided, despite the fact that, as in the case of earthquakes, a
scale-free avalanche event dynamics (e.g. the Gutenberg-Richter law)
can be observed in many far from equilibrium phenomena which exhibit
$1/f^{\alpha}$ noise (and non-Poissonian inter-event statistics
\cite{davidsen01a,bak-earth}) at vastly longer time scales (e.g. days
to years) compared to the individual events (e.g. seconds).  Perhaps
the lack of a sufficiently convincing ``sandpile'' model  has
contributed to the ongoing debate about the relevance of SOC to solar
flares, earthquakes, transport in tokamaks, and many other burst-like
phenomena.

The ``sandpile'' model we discuss  was introduced by de Sousa Vieira to
describe avalanches in stick-slip phenomena \cite{vieira00}.  It is close to
the original array of connected pendula first discussed by Bak, Tang,
and Wiesenfeld.  In detail, the model is defined as follows: Consider a
one-dimensional system of size $L$ where a continuous variable $f_n
\geq 0$ is associated with each site $n$ (representing the force on
that site).  Initially, all $f_n$ have the same value which is below a
threshold $f_{th}$.  The basic time step of the driving force consists
in changing the value of the force on the first site according to $f_1
= f_{th} + \delta f$ with a fixed $\delta f$.  This can be considered
as a slow external driving and leads to a fast relaxation process
(avalanche) within the system.  This relaxation consists of a
conservative redistribution of the force at sites with $f_n \geq
f_{th}$ (toppling sites)
 according to $f_n = \Phi(f_n - f_{th})$ and $f_{n \pm 1} =
f_{n \pm 1} + \Delta f_n/2$.  Here, $\Delta f_n$ is the change of
force at the overcritical site and $\Phi$ a periodic, nonlinear
function.  The relaxation continues until all sites are stable again,
i.e., $f_n < f_{th}$ for all $n$.  Then, the driving at the first site
sets in again.  This definition is complemented by open boundary
conditions, i.e., force is lost at both boundaries.  Without loss of
generality, we use a sequential update and set $f_{th} = 1$, $\delta f
= 0.1$, $\Phi (x) = 1 - a [ x ]$ where $[ x ]$ denotes $x$ modulo
$1/a$, i.e., a sawtooth function, and $a = 4$.  It was shown that the
model evolves into a state of SOC where the avalanche distributions
are scale free, limited only by the overall system size
\cite{vieira00}.  Note that like other sandpile models,
 the toppling rules  are conservative,
and the total amount of force in the system can only be changed
at the boundaries.

\begin{figure}[bt]
\centerline{\psfig{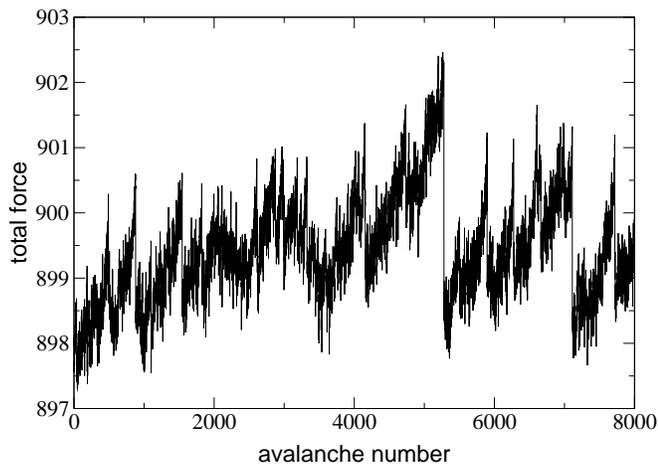}}
\vspace{0.25cm}
\caption{Fluctuations of the total force in the self-organized critical state 
for a system of size $L=1024$.}
\label{signal}
\end{figure} 
An
appropriate choice of a time signal to detect $1/f^{\alpha}$ noise, which takes
the time scale separation between the slow external driving and the 
individual avalanches into account, is the total force in the system after
each avalanche:
\begin{equation}
X(i) = \sum_{n=1}^{L} f_n(i),
\label{xt}
\end{equation}
where $i$ is the avalanche number. This signal is shown in
Fig. \ref{signal} and directly reveals the stick-slip character of the
dynamics.

\begin{figure}[bt]
\centerline{\psfig{figure=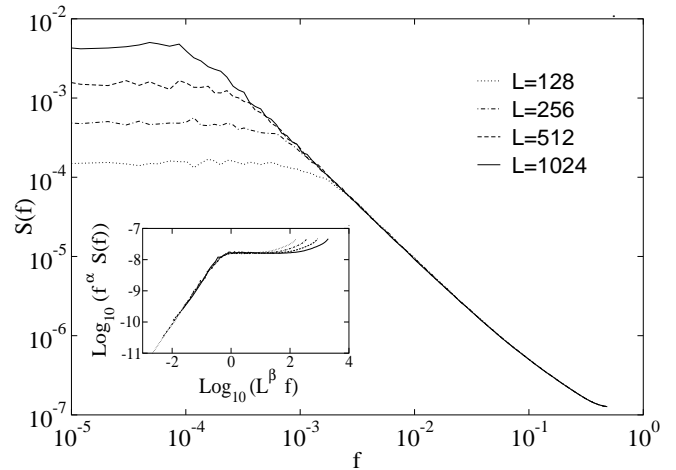,width=\columnwidth,clip=}}
\vspace{0.25cm}
\caption{Power spectrum of the fluctuations in the total force for different 
system sizes. Note that the cut off moves to lower frequencies for larger 
system sizes. Inset: Data collapse with $\alpha = 1.38$ and $\beta = 1.2$.}
\label{ps}
\end{figure}

Analyzing the power spectrum of $X(i)$, we find a clear $1/f^\alpha$
decay with a cut off at low frequencies that shifts to even lower
frequencies with increasing system size (see Fig. \ref{ps}).  In
particular, a data collapse, shown in the inset, reveals
the following scaling behavior:
\begin{equation}
S(f) \sim \frac{1}{f^\alpha} g(\frac{1}{f L^\beta}),
\end{equation}
with $\alpha = 1.38$ and $\beta = 1.2$.  Here, $g$ is a scaling
function that is constant for small arguments and decays as
$x^{-\alpha}$ for large arguments.  These results are
somewhat dependent on the details of the definition of the model
\cite{foot1}, a fact which does not change our main conclusion regarding
SOC as a mechanism leading to $1/f^{\alpha}$ noise.  

To understand how the long range temporal correlations arise in this model,
we have
analyzed the fluctuations in the force at a single site, i.e.
$X_n(i) = f_n(i)$.  It turns out that the local power
spectra are simple Lorentzians with a characteristic frequency that
decreases with increasing distance from the driving site. 
Thus, different time signals of the
same system, such as the local and the total force signal, behave
in a totally different manner, confirming  earlier results 
\cite{kogan}.  Moreover, the observation of
Lorentzians suggests that the $1/f^{\alpha}$ noise could occur due
to a superposition of independent and purely random processes each
described by a characteristic time scale $t_c$ (see, e.g., 
Ref. \cite{kogan}).
However, the fluctuations of the force at different
sites are correlated.  This can already be deduced from the fact that
the force at a given site can only change if all sites closer to the
first site have discharged as well.  A more careful analysis shows
that the assumption of uncorrelated (local) signals would give a value
of $\alpha$ far away from the observed value
\cite{davidsen_up}.
Consequently, the long range temporal
correlations, implied by the occurrence of $1/f^\alpha$ noise, are
stored in the spatial correlations embedded in the whole system.

This is further confirmed by the fact that scaling relations connect $\alpha$
and $\beta$ to the critical exponent characterizing the avalanche
distributions.  The quantity $L^{\beta}$ in Eq. (2) describes the 
scaling of the temporal cutoff and, hence, the scaling of the number of
avalanches before events become uncorrelated. Thus,  it can be related
to the avalanche distribution. 
The distribution of avalanche sizes (the
number of toppling events in the avalanche) is distributed as $P(s)
\sim s^{-\tau}G(s/L^D)$, where $\tau \simeq 1.54$ is the so-called
histogram exponent, and the avalanche dimension $D\simeq 2.20$ gives
the cutoff for the largest avalanche in a system of size $L$ \cite{vieira00}.  
Avalanches
in the power-law regime do not extend through the entire system; therefore
some sites do not topple and the system retains memory of its
previous force.  However, avalanches larger than 
 the cutoff, which are the ones that
entirely span the chain, decorrelate the system because all
sites topple and get a new force.  The frequency of
the large system wide events that decorrelate the force in the
system scales as $L^{-D(\tau -1)}$. Thus, $\beta = D(\tau -1)$.
Also, a scaling relation $\langle s \rangle \sim L$  was found 
numerically in \cite{vieira00}, and is a general result for many boundary
driven SOC systems \cite{pb}, giving the result that $D(2-\tau)=1$.
Combining the two equations, $\beta=D-1=(\tau -1)/(2-\tau)$, 
which agrees very well with numerical results.

Using scaling relations together with some previously obtained results
concerning universality classes,
we can relate the exponent $\alpha$ to the exponent $\tau$ or $D$ via
the variance in the total force in the system:
\begin{equation}
\int S(f) df = \sigma^2 = \left \langle \left (\sum_{n=1}^{L} f_n(i) - \left \langle
\sum_{n=1}^{L} f_n(i) \right \rangle \right)^2 \right \rangle,
\end{equation}
where the integral is over all frequencies.  Since the exponent
$\alpha >1$, the power in the signal diverges as $L \rightarrow
\infty$ as $L^{\beta(\alpha -1)}$.  As argued and supported
numerically in Ref. \cite{vieira00}, the ``sandpile'' model is in
the same universality class as the original Burridge-Knopoff train
model.  This latter model has been conjectured to be in the universality
class of interface depinning with the interface pulled at one end
\cite{pb} -- a result which is also supported by numerical
simulations.  Thus, the fluctuations in the force in the system have
the dimension of fluctuations of force in the interface depinning
problem, e.g. $(\int_0^L dx \nabla^2 H(x))^2$ (see Ref. \cite{pb} for
more explanation).  Since in that case, the height of the interface
$H$ has the dimension of $L^{D-1}$, then $\sigma^2 \sim L^{2(D-2)}$.
Using the previous relation for $\beta$, we get $\alpha=(3D-5)/(D-1)=
(5\tau -7)/(\tau -1)$.  This again agrees very well with our numerical
simulation results and is consistent with the results in
\cite{zhang00}.
The existence of these scaling relations makes it clear without
ambiguity that, in this model, the critical avalanche dynamics and the 
long range temporal correlations belong inseparably together.

\begin{figure}[bt]
\centerline{\psfig{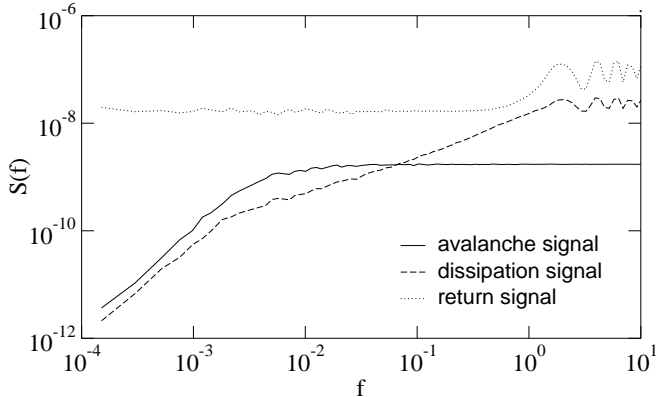}}
\vspace{0.25cm}
\caption{Power spectrum of the different, differentiated, time signals
for $L=256$ (see text). Note that in the avalanche and dissipation
signal, an increase with $f^2$ as observed for the lowest frequencies
corresponds to white noise in the integrated signal. The avalanche
signal is shifted down by 8 decades.}
\label{aps}
\end{figure}

Instead of using the 
avalanche number,  we can choose different definitions for
the slow time scale of the model.  Since
the force on the first site is set to $f_{th} + \delta f$ to induce a
new avalanche,
one can choose to identify the amount of force added to the
first site with the temporal interval between avalanches. This
corresponds to a uniform driving.
Such a choice leads
to non-uniform time intervals between successive relaxation events 
which could in principle alter the behavior.
The time signal defined in
Eq. (\ref{xt}) is modified and becomes a sawtooth function
\begin{equation}
X(t) = \sum_{n=1}^{L} f_n(t) = t - t_i + \sum_{n=1}^{L} f_n(t_i),
\label{xc}
\end{equation}
where $t_i \leq t$ denotes the time of occurrence of the last
avalanche.  The power spectrum of this time signal shows exactly the
same $1/f^{\alpha}$ decay as the signal defined in Eq. (\ref{xt}) as
can be deduced from Fig. \ref{aps}.  There, we show the power spectrum
of the differentiated signal $X'(t)$ --- called a dissipation signal for
obvious reasons --- which follows a power law with $S(f) \propto
f^{2-\alpha}$. In particular, the cut off to white noise behavior 
$(\alpha =0)$ for
very low frequencies occurs almost at the same frequency as before and
the scaling with $L$ is also unchanged. Note that white noise,
$(S(f) \sim f^0)$ in $X(t)$ in Fig. 2, corresponds to $S(f) \propto f^2$ in
$X'(t)$ in Fig. 3.

The dissipation signal resembles the form of a pulse train, i.e.,
\begin{equation}
X'(t) = \sum_i h_i \delta (t-t_i),
\label{diffsig}
\end{equation}
where $h_i$ is the dissipation in the total force due to the $i$th
avalanche.  This form of the signal is especially well-suited to 
clarify the source of the long time correlations.  For
instance, substituting $h_i$ by a constant should reveal the
correlations induced solely by the fluctuations in the time intervals
between dissipation events.  As
shown in \cite{davidsen01a,kaulakys98}, correlations
between subsequent time intervals can, indeed, lead to $1/f^\alpha$ noise.
However,
here we find that this ``return signal'' is uncorrelated
(see Fig. \ref{aps}). The absence of correlations
in the time intervals together with the existence of a finite second 
moment of their distribution due to the boundedness of the intervals 
by $f_{th} + \delta f$ implies that the power spectrum is the same as
before for frequencies below a certain, fixed $f_h$. 
Hence, the long time scales are not affected by the fluctuations in 
the time intervals.
In particular, this is true for any distribution of time intervals
with finite second moment if the time intervals are independent of 
one another (central limit theorem).
Thus, the $1/f^{\alpha}$ noise is due solely to the fluctuations in $h_i$.
We conclude that the long range temporal correlations
in the  model are exclusively encoded in the state of the whole system
and are not destroyed by the fluctuations due to changes
in the definition of the time scale of the external driving.

The fact that the occurrence of $1/f^{\alpha}$ noise in the 
model cannot be attributed to a low-dimensional dynamics shows that
the SOC mechanism is totally different and can be well-distinguished
from other deterministic mechanisms which attempt to explain flicker
noise as a chaotic phenomenon (see, e.g.,
Refs. \cite{schuster}).  Unlike those mechanisms, the
dynamics discussed here cannot be reduced to a renewal process with a
power-law distribution of waiting times or step sizes.

Finally, the temporal variations in the
sizes of the avalanches show only trivial long time correlations.
Fig. \ref{aps} shows the power spectrum of the avalanche signal,
defined in (\ref{diffsig}) where $h_i$ is now the number of topplings
in  the $i$th
avalanche. Clearly, a flat spectrum can be observed with a change to
$f^2$ behavior at
low frequencies due to the finite system size. In terms of the 
integrated signal $X(t)$, this corresponds to a trivial $1/f^2$ decay 
crossing over to 
white noise $(f^0)$  behavior at low frequencies.  Hence, the underlying 
process can be considered as a random walk confined by the system
size.  

To summarize, the interplay between building up of force
from the external driving when the avalanches are small, balanced by
dissipation of force at the boundaries when the avalanches are larger is
responsible for the flicker noise in this SOC  model.
Our results explicitly demonstrate
 that a SOC mechanism, with a power-law distribution of avalanches, 
can lead to long time correlations between avalanches
 and thus low frequency $1/f^{\alpha}$ noise in slowly driven
systems.

We thank P. Bak for useful discussions.
J. D. would like to thank the DAAD for financial support and the Imperial
College for support and hospitality.

% figures follow here
%
% Here is an example of the general form of a figure:
% Fill in the caption in the braces of the \caption{} command. Put the label
% that you will use with \ref{} command in the braces of the \label{} command.
%


\begin{references}
\bibitem{foot1}
The same exponents are observed if the boundary at the last site is closed.
Moving the driving site to the center of the chain or closing the boundary
at the driving site lead to a totally different behavior.  Although the 
avalanche distribution is still scale free, 
the $1/f{^\alpha}$ noise is destroyed.
\end{references}
\end{document}